\documentclass[showpacs,prl,twocolumn]{revtex4}
\usepackage{graphicx}
\usepackage{dcolumn}
\usepackage{bm}
\begin{document}

\title{Signatures of Magnetized Large Scale Structure in Ultra-High
Energy Cosmic Rays}
\author{G\"unter Sigl$^a$, Francesco Miniati$^b$,
Torsten~A.~En\ss lin$^b$}
\affiliation{$^a$ GReCO, Institut d'Astrophysique de Paris, C.N.R.S.,
98 bis boulevard Arago, F-75014 Paris, France\\
$^b$ Max-Planck Institut f\"ur Astrophysik,
Karl-Schwarzschild-Str.~1, 85741 Garching, Germany}

\begin{abstract}
We investigate the impact of a structured universe in the multi-pole
moments, auto-correlation function, and cluster statistics
of cosmic rays above $10^{19}\,$eV. We compare
structured and uniform source distributions with and without
magnetic fields obtained from a cosmological
simulation. We find that current data marginally favor 
structured source distributions and magnetic fields
reaching a few $\mu$G in galaxy clusters but below $0.1\,\mu$G
in our local extragalactic neighborhood.
A pronounced GZK cutoff is also predicted in this scenario.
Future experiments will make the degree scale auto-correlation
function a sensitive probe of micro Gauss fields surrounding the sources.
\end{abstract}

\pacs{98.70.Sa, 13.85.Tp, 98.65.Dx, 98.54.Cm}

\maketitle

{\it Introduction:}
The origin and nature of ultra high energy cosmic rays (UHECRs)
is one of the most puzzling problems of high energy 
astrophysics~\cite{bs-rev}.
Beside the actual mechanism responsible for the acceleration of 
the particles to energies as high as a few $\times 10^{20}\,$eV 
and more, propagation of UHECRs in intergalactic space is a delicate
and critical part of the full investigation. 
It was soon recognized that energy losses dramatically limit the
possible distance traveled by these particles; in particular,
interactions with cosmic microwave background photons lead
to catastrophic photo-meson production at energies above
$\sim 5\times10^{19}\,$eV, the GZK effect~\cite{gzk}.
Charged nucleons are also susceptible to deflections induced by
extragalactic magnetic fields (EGMF). 
The importance and the subtlety of the effects 
produced by this process have been emphasized by a number of
authors~\cite{sigletal,sse,ynts,tanco}. However, up to date, the
properties of EGMF are only very poorly constrained;
therefore, their impact 
on the propagation of UHECRs is still quite uncertain and a number of 
different scenarios, compatible with the available observational 
constraints, must be investigated.

Along this line of reasoning, recently we have investigated UHECR
propagation in a highly structured EGMF obtained from a
simulation of large scale structure formation~\cite{sme}. The
EGMF was evolved according to the induction equation. Its
seeds were generated at shocks in accord to the Biermann battery
mechanism~\cite{kcor97}.  Thus, it was amplified in different parts of
the universe according to the velocity field provided by the simulated
gas component. Lorentz forces were neglected as EGMF are
known to be dynamically unimportant in most of the universe
volume. The resulting field strength was normalized so as to reproduce
observed rotation measure 
values in galaxy clusters~\cite{bo_review}. The EGMF
are significant only within filaments and groups/clusters of
galaxies. The relative strength there and the resulting overall
topology depend solely on hydrodynamics which is basically driven by
gravitational instability.  About 90 percent of the volume is filled
with negligible fields.  This is a reflection of the adopted
Biermann field generation scheme, as opposed to the case in which the
initial magnetic field is set uniform over the whole volume.

Here we extend our previous study~\cite{sme} to include sources
at cosmological distances, model more realistically the UHECR source
distributions and compare explicitly with the case of no EGMF.
Our favored scenario is in considerably improved agreement with UHECR 
data and reveals signatures of the EGMF to which current cosmic ray 
data start to become sensitive.

{\it Numerical Techniques:}
The simulation of large scale structure formation was carried out
within a volume of size $50\,h^{-1}\,$Mpc, on a side, where
$h\equiv H_0/(100$ km s$^{-1}$ Mpc$^{-1})$ = 0.67, using
a comoving grid of 512$^3$ zones and 256$^3$ dark matter
particles~\cite{miniati}. Two observer positions are compared as in
Ref.~\cite{sme}: The first in a filament-like structure with EGMF
$\sim0.1\,\mu$G and the second at the border of a small void with
EGMF $\sim10^{-11}\,$G. A relatively large structure about 17 Mpc away
from the weak field observer is identified for calculation purposes 
as the Virgo cluster. We orient our terrestrial coordinate system so
that this cluster is close to the equatorial plane.

For a given number density of UHECR sources, $n_s$, 
we explore both the case in which their spatial distribution is 
either proportional to the local baryon density, as in Ref.~\cite{sme}, or
completely uniform.
Furthermore, we assume that these sources are distributed (1) with 
respect to their (dimensionless) UHECR emission power, $Q_i$, as
$\partial n_s/\partial Q_i\propto Q_i^{-2.2}$ with $1\leq Q_i\leq100$,
as motivated by the luminosity function of the EGRET $\gamma-$ray blazars
in the power range $10^{46}\,{\rm erg}\,{\rm s}^{-1}\lesssim Q_i
\lesssim10^{48}\,{\rm erg}\,{\rm s}^{-1}$~\cite{cm}.
And (2) with respect to the spectral index $\alpha_i$ of the emitted power-law
distributions of UCHRs as $\partial n_s/\partial\alpha_i=$const for
$\alpha-0.1\leq\alpha_i\leq\alpha+0.1$, where $\alpha ~(\sim 2.5)$ is
a free parameter to be determined through a best fit analysis. Each
source accelerates up to $10^{21}\,$eV, and its average UHECR emission
characteristics do not change significantly on the time scale of UHECR
propagation.  Injected power and spectral index are then fit to
reproduce the observed spectrum.

Taking advantage of the periodic boundary conditions, 
periodic images of the simulation box are added until the linear
size of the enclosed volume is larger than
the energy loss length of nucleons above $10^{19}\,$eV, $\sim1\,$Gpc.
As in previous work ~\cite{sigletal,sme}, particles are propagated
taking into account Lorentz forces due to the EGMF and energy losses.
In this respect, pion production is treated stochastically,
whereas pair production as a continuous energy loss process.
Trajectories connecting sources and observers in different copies
of the simulation box are taken into account.

An {\it event} was registered and its arrival directions and 
energies recorded each time the trajectory of the propagating particle
crossed a sphere of radius 1.5 Mpc around one of the observers.
For each configuration $10^4$ events (a realization) where collected to
construct simulated sky distributions. These are multiplied
with the solid-angle dependent exposure function and folded over the
angular resolution characteristic of a given experiment. 
Finally from each simulated sky distribution typically 100 mock 
data sets, consisting of $N_{\rm obs}$ observed events, 
were randomly extracted. The angular resolution is 
$\simeq1.6^\circ$ above $4\times10^{19}\,$eV and
$\simeq2.5^\circ$ above $10^{19}\,$eV for the AGASA
experiment~\cite{agasa} and $\simeq10^\circ$ for the 
SUGAR data~\cite{sugar}. For the exposure function we use the
parameterization of Ref.~\cite{sommers} with parameters as in
Ref.~\cite{anchor_iso}. Energy resolution effects, supposed to be of
order $\Delta E/E\simeq30\%$, are also taken into account.

For each data set (both simulated and real) we obtained estimates
for the spherical harmonic coefficients $C(l)$, the autocorrelation
function $N(\theta)$, and the number of multiplets $M(n)$ of $n$
events within an angle $\theta_m$ via the following estimator functions
$C(l)=(2l+1)^{-1}\sum_{m=-l}^l\left(\sum_{i=1}^NY_{lm}(u^i)/\omega_i/
\sum_{i=1}^{N}1/\omega_i\right)^2$, where $\omega_i$ is the total
experimental exposure at arrival direction $u^i$, and $Y_{lm}(u^i)$
is the real-valued spherical harmonics function taken at direction
$u^i$~\cite{sme}; 
$N(\theta)$ is defined as $N(\theta)\propto\sum_{j \neq i}\left\{
1\,\mbox{if $\theta_{ij}$ is in same bin as $\theta$, 0 otherwise}
\right\}/S(\theta)$, where $S(\theta)$ is the solid angle size of
the corresponding bin, and normalized to one for an isotropic
distribution.

The mock data sets in the various realizations yield the statistical
distributions of $C(l)$, $N(\theta)$, and $M(n)$.
We define the average over all mock data sets and realizations
and two errors. A smaller, statistical
error, i.e. the fluctuations due to the finite number $N_{obs}$ of observed
events, averaged over all realizations. And a larger,
``total error'', i.e. the statistical error plus the
cosmic variance due to the variation in source characteristics.
Typically, for each scenario we consider 10
realizations for the source positions for each of which we chose 50
different realizations for the power $Q_i$, and injection spectral index
$\alpha_i$, sampled from the distributions discussed above.

Given a set of observed and simulated events, we define
$\chi_4\equiv\sum_i\left[\left(S_{i,{\rm data}}-
\overline{S}_{i,{\rm simu}}\right)/\Delta S_{i,{\rm simu}}\right]^4$,
where $S_i$ stands for $C_l$, $N(\theta)$, or $M(n)$,
$S_{i,{\rm data}}$ refers to $S_i$ obtained from the real data, and
$\overline{S}_{i,{\rm simu}}$ and $\Delta S_{i,{\rm simu}}$ are the
average and standard deviations of the simulated data sets.  This
measure of deviation from the average prediction can be used to
obtain an overall likelihood for the consistency of a given
theoretical model with an observed data set by counting the fraction
of simulated data sets with $\chi_4$ larger than the one for the
real data. For further details see Refs.~\cite{sigletal,sme,miniati,upcoming}.

{\it Constraints from Existing Data and Predictions for Future Experiments:}
The scenarios studied and results are summarized in Tab.~\ref{tab1}.
UHECR sources whose number density is given in column II
are distributed either proportionally to
the simulated baryonic density or uniformly (``yes'' or ``no'' respectively 
in column III). Finally, the EGMF is either
taken from the simulation with local value as indicated in column IV,
or completely neglected (``no EGMF'').
The number of realizations of source positions was 10 except for
scenario 1 for which it is 7.

\begin{table*}[ht]
\caption[...]{List of simulated scenarios. The columns contain
the number assigned to the scenario, the source density, whether the
sources are distributed as the baryon density in the
simulation box or uniformly (yes/no), the magnetic field
strength at the observer location, the best fit power
law index in the injection spectrum $E^{-\alpha}$, and the overall likelihoods
of fits to the data. The first four likelihoods are for the
multi-poles above the energy indicated as superscript
in EeV and over the range of $l$ indicated as subscript. Above 40 EeV
the comparison was made with $N_{\rm obs}=99$ AGASA+SUGAR events, whereas
above 10 EeV comparison with an isotropic distribution of 1500 events
was made, see text. The last two likelihoods are for the auto-correlation
for $\theta\leq20^\circ$, and the clustering within
$2.5^\circ$ up to multiplicity 10, respectively.}\label{tab1}
\begin{ruledtabular}
\begin{tabular}{ccccccccccc}
\#&$n_s$ [Mpc$^{-3}$]& structure & $B_{\rm obs}/$G &
$\alpha$ &${\cal L}^{40}_{l\leq10}$&
${\cal L}^{40}_{l=1}$&${\cal L}^{10}_{l\leq10}$&${\cal L}^{10}_{l=1}$&
${\cal L}^{40}_{\theta\leq20^\circ}$&${\cal L}^{40}_{n\leq10}$\\
\hline \\
1 &$2.4\times10^{-4}$& yes & $1.3\times10^{-7}$ & 2.4 & 0.34
& 0.087 & 0.13 & 0.035 & 0.56 & 0.88 \\
2 &$2.4\times10^{-4}$& yes & $8.2\times10^{-12}$ & 2.4 & 0.52
& 0.48 & 0.16 & 0.18 & 0.52 & 0.85\\
3 &$2.4\times10^{-4}$& yes & no EGMF & 2.6 & 0.37
& 0.39 & 0.15 & 0.15 & 0.42 & 0.73 \\
4 &$2.4\times10^{-5}$& yes & no EGMF & 2.6 & 0.33
& 0.48 & 0.13 & 0.19 & 0.30 & 0.65 \\
5 &$2.4\times10^{-5}$& no & no EGMF & 2.6 & 0.45
& 0.51 & 0.15 & 0.22 & 0.65 & 0.71 \\
6 &$2.4\times10^{-5}$& yes & $8.2\times10^{-12}$ & 2.4 & 0.79
& 0.62 & 0.17 & 0.24 & 0.56 & 0.83 \\
\end{tabular}
\end{ruledtabular}
\end{table*}

\begin{figure}[ht]
\includegraphics[width=0.49\textwidth,clip=true]{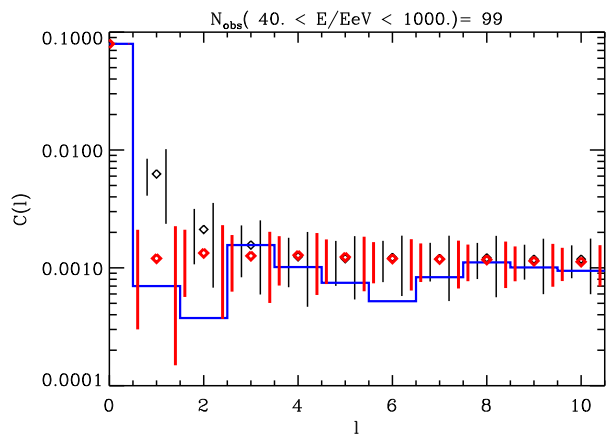}
\caption[...]{Angular power spectrum $C(l)$ as a function of
multi-pole $l$, obtained for the AGASA+SUGAR exposure function,
for $N_{\rm obs}=99$ events observed above 40 EeV. We show 
the realization averages (diamonds), statistical (left) and total (right)
error bars, respectively, predicted by the model. 
The red (thick) diamonds and red (thick, outer)
error bars represent scenario 6, whereas the black (thin) diamonds
and black (thin, inner) error bars represent scenario 1.
The histogram corresponds to the AGASA+SUGAR data.}
\label{fig1}
\end{figure}

It was noted in Ref.~\cite{anchor_iso} that the AGASA and SUGAR
experiments had comparable exposure in the northern and southern
hemisphere, respectively. Above $4\times10^{19}\,$eV they combined
50 events observed by AGASA (excluding 7 events observed by Akeno)
and 49 events seen by SUGAR. While SUGAR's angular resolution is
much worse than for AGASA and in general prevents a combination of
the two data sets, multi-poles $l\leq10$ are not sensitive to scales
$\lesssim10^\circ$. To take advantage of more
statistics and more complete sky coverage, we therefore use
the combined AGASA+SUGAR data set of 99 events above $4\times10^{19}\,$eV
when computing multi-poles $l\leq10$. The results for scenarios 1 and 6 in
Tab.~\ref{tab1} are compared in Fig.~\ref{fig1}. In contrast, since
the auto-correlation
function and clustering are sensitive to small scales, when
comparing with data above $4\times10^{19}\,$eV we will
only use the reported 57 AGASA+Akeno events.

\begin{figure}[ht]
\includegraphics[width=0.49\textwidth,clip=true]{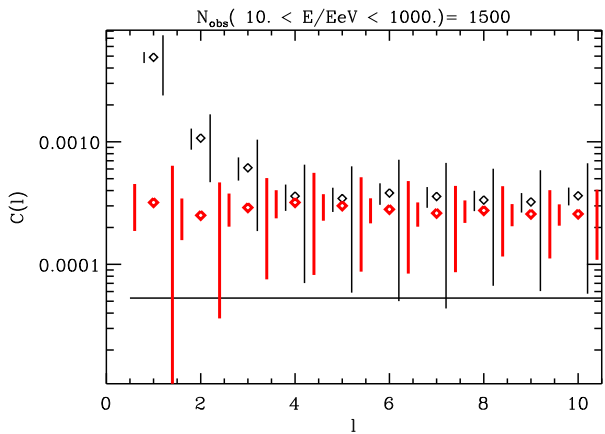}
\caption[...]{As Fig.~\ref{fig1}, but for the angular power spectrum
obtained for the combined AGASA+SUGAR exposure function,
for $N_{\rm obs}=1500$ events observed above 10 EeV. The straight
line is the analytical prediction, $C_l\simeq(4\pi N)^{-1}$, for the
average multi-poles for complete isotropy.}
\label{fig2}
\end{figure}

No signs of anisotropy were found by either the AGASA or the SUGAR
experiment down to $10^{19}\,$eV. We therefore also compute predictions
for the multi-poles for $\simeq1500$ events observed above $10^{19}\,$eV,
using the combined exposure of AGASA+SUGAR. This is compared with an
isotropic distribution in Fig.~\ref{fig2} for scenarios 1 and 6.
The likelihoods for consistency with isotropy are also summarized in
Tab.~\ref{tab1}.

\begin{figure}[ht]
\includegraphics[width=0.49\textwidth,clip=true]{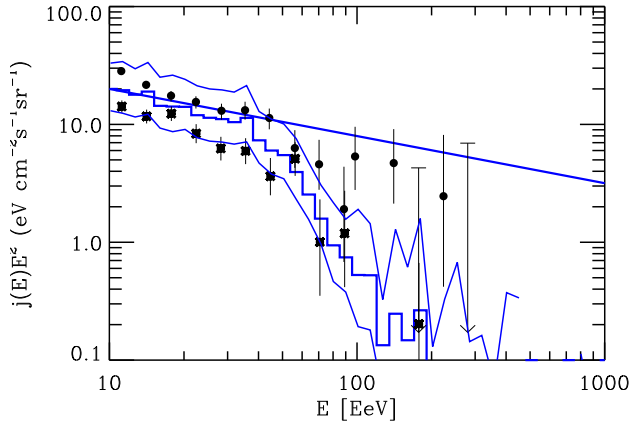}
\caption[...]{Predicted spectrum observable by AGASA for scenario
6, for which multi-poles were shown in Fig.~\ref{fig1}, as compared to the
AGASA (dots) and HiRes-I~\cite{hires} (stars) data. The histogram marks the
average and the two lines above and below the 1-sigma fluctuations
over the simulated realizations. The solid
straight line marks the injection spectrum.}
\label{fig3}
\end{figure}

Comparison in Tab.~\ref{tab1} of multi-poles, auto-correlation function
and clustering above $4\times10^{19}\,$eV, as well as observed
isotropy at $10^{19}\,$eV, see Fig.~\ref{fig2},
favor an observer immersed in relatively weak
fields, especially scenario 6. For the multi-poles
describing the large scale distribution this can qualitatively be
understood as follows: Relatively strong fields surrounding the
observer suppress the flux from the far sources. The observed flux
is therefore more strongly dominated by the nearest few sources and
thus more anisotropic. For observers immersed in weak fields,
a relatively strong GZK cutoff is expected, inconsistent with AGASA
data, see Fig.~\ref{fig3}, and in contrast to the disfavored strong
field observer.

\begin{figure}[ht]
\includegraphics[width=0.49\textwidth,clip=true]{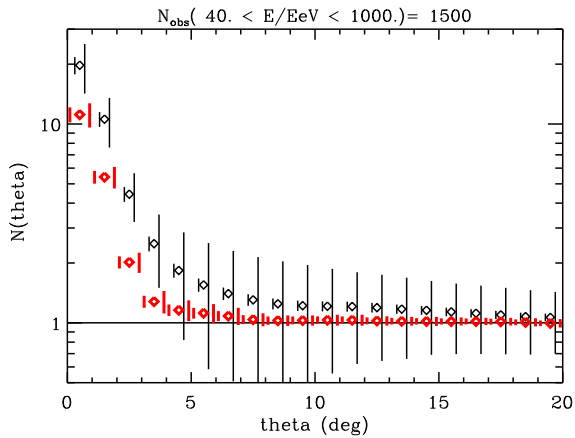}
\caption[...]{The auto-correlation function $N(\theta)$ as a function
of angular distance $\theta$ for $N_{\rm obs}=1500$ events
observed above 40 EeV in $1^\circ$ bins. The key for the model predictions
is as in Figs.~\ref{fig1} and~\ref{fig3}, now comparing scenario
6 (lower set) and 4 (upper set) instead, which only differ in the EGMF.
$N(\theta)=1$ corresponds to isotropy.}
\label{fig4}
\end{figure}

Scenarios 3 and 4 are somewhat disfavored by auto-correlation
and clustering: Structured sources produce more clustering in the
absence of magnetic fields. While this effect is only marginal in 
the current data set, it will provide a solid criterion
for discriminating between the two scenarios in future
experiments: Fig.~\ref{fig4} shows that the auto-correlation
function is sensitive to the magnetic fields around the sources.
Scenarios with no significant magnetic fields predict a stronger
auto-correlation at small angles, independent of whether or not
the sources are structured. This is because the images of sources
immersed in considerable magnetic fields are smeared out,
which also smears out the auto-correlation function over a few
degrees. As shown in Fig.~\ref{fig4}, this effect will become
quite significant once $\sim1500$ events are observed above
$4\times10^{19}\,$eV, for example, by the Pierre Auger project~\cite{auger}.
It also appears robust against variation of other parameters such
as source density and distribution: For example, whereas
$N(1^\circ)\simeq10.1\pm1.6$
for scenario 6, all other scenarios without magnetic field predict
$N(1^\circ)\gtrsim16\pm3$.
Finally, all scenarios considered predict significant anisotropy
above $4\times10^{19}\,$eV for exposures reached by the Pierre Auger
project.

{\it Conclusions:}
Our results in Ref.~\cite{sme} showed that, in the case of structured
extragalactic magnetic fields, the data were marginally consistent with 
an observer immersed in $\sim0.1\,\mu$G field, although 
the isotropy of UHECRs at $10^{19}\,$eV
could not be explained by a local source distribution alone. 
In the present paper we took into account sources up to cosmological
distances and adopted more realistic distributions for the emission
properties of UHECR sources.
This considerably improved the quality of our fits at all energies. 
We conclude that the {\it combination} of statistical quantities 
characterizing the comparison between
observed and simulated data moderately favors a scenario in which 
(i) UHECR sources have a density $n_s\sim10^{-5}\,{\rm Mpc}^{-3}$
and follow the matter distribution
(ii) magnetic fields are relatively pervasive within the large
scale structure, including filaments, and with a strength of order of
a $\mu$G in galaxy clusters (iii) the local extragalactic environment 
is characterized by a weak magnetic field below $ 0.1\,\mu$G.
Thus, current UHECR data already allow us to probe the large
scale distribution of matter and magnetic fields
and seem to marginally confirm the quite realistic paradigm summarized 
above!
This will increasingly be true for
future detectors such as the Pierre Auger~\cite{auger} and EUSO~\cite{euso}
projects. In particular,
the degree-scale auto-correlation functions above
$\simeq4\times10^{19}\,$eV can serve as a discriminator between
magnetized and unmagnetized sources: $N(1^\circ)\lesssim10$ points
to strong magnetization, rather independently of other source parameters.
The best fit scenario also predicts a pronounced GZK cutoff.

FM acknowledges support by the Research and 
Training Network ``The Physics of the Intergalactic Medium''
under EU contract HPRN-CT2000-00126 RG29185.
GS thanks the Max-Planck-Institut f\"ur Physik
for hospitality and financial support.


\begin{thebibliography}{99}

\bibitem{bs-rev} P.~Bhattacharjee and G.~Sigl,
Phys.~Rept. 327 (2000) 109; L.~Anchordoqui, T.~Paul, S.~Reucroft,
and J.~Swain, Int.~J.~Mod.~Phys. A18 (2003) 2229.

\bibitem{gzk} K.~Greisen, Phys.~Rev.~Lett. 16 (1966)
748; G.~T.~Zatsepin and V.~A.~Kuzmin, Pis'ma
Zh. Eksp. Teor. Fiz. 4 (1966) 114 [JETP. Lett. 4 (1966) 78].

\bibitem{sigletal} G.~Sigl, M.~Lemoine, and P.~Biermann,
Astropart.~Phys. 10 (1999) 141; M.~Lemoine, G.~Sigl, and P.~Biermann,
e-print astro-ph/9903124; C.~Isola, M.~Lemoine, and G.~Sigl,
Phys.~Rev.~D 65 (2002) 023004; C.~Isola and G.~Sigl, Phys.~Rev.~D 66
(2002) 083002.

\bibitem{sse} T.~Stanev et al., Phys.~Rev.~D 62 (2000) 093005.

\bibitem{ynts} H.~Yoshiguchi, S.~Nagataki, S.~Tsubaki, and K.~Sato,
Astrophys.~J. 586 (2003) 1211; H.~Yoshiguchi, S.~Nagataki, and K.~Sato,
Astrophys.~J. 592 (2003) 311; astro-ph/0307038.

\bibitem{tanco} G.~Medina Tanco, Lect.~Notes~Phys. 576 (2001) 155.

\bibitem{sme} G.~Sigl, F.~Miniati, and T.~En\ss lin, Phys.~Rev.~D 68
(2003) 043002.

\bibitem{kcor97} R. M. Kulsrud, R. Cen, J. P. Ostriker, and D. Ryu,
Astrophys.~J., 480 (1997) 481

\bibitem{bo_review} for a recent review see, e.g., J.-L.~Han
and R.~Wielebinski, CHJA\&A 2 (2002) 293 [e-print astro-ph/0209090].

\bibitem{miniati} F.~Miniati, Mon.~Not.~R.~Astron.~Soc. 337 (2002) 199.

\bibitem{cm} J.~Chiang and R.~Mukherjee, Astrophys.~J. 496 (1998) 752.

\bibitem{agasa} M.~Takeda et al., Phys. Rev. Lett. 81 (1998) 1163;
Astrophys. J. 522 (1999) 225; Hayashida et al.,
e-print astro-ph/0008102; see also
{\sf http~://www-akeno.icrr.u-tokyo.ac.jp/AGASA/}.

\bibitem{sugar} M.~M.~Winn et al., J.~Phys.~G 12 (1986) 653; see also
{\sf http://www.physics.usyd.edu.au/hienergy/sugar.html}.

\bibitem{sommers} P.~Sommers, Astropart.~Phys. 14 (2001) 271.

\bibitem{anchor_iso} L.~Anchordoqui et al., e-print astro-ph/0305158.

\bibitem{upcoming} G.~Sigl, F.~Miniati, and T.~En\ss lin, in preparation.

\bibitem{hires} T.~Abu-Zayyad et al. (HiRes collaboration),
e-print astro-ph/0208243; e-print astro-ph/0208301.

\bibitem{auger} J.~W.~Cronin, Nucl.~Phys.~B (Proc.~Suppl.) 28B (1992)
213; The Pierre Auger Observatory Design Report (ed.~2), March 1997;
see also {\sf http://www.auger.org}.

\bibitem{euso} See {\sf http://www.euso-mission.org}.

\end{thebibliography}
\end{document}